\documentclass[12pt,preprint,a4paper]{aastex}

\usepackage{graphics,epsf}
\usepackage{amsmath}                
\usepackage{amsfonts}               
\usepackage{amssymb}                
\usepackage{epsfig}                 

\def \cm{~\rm{cm}}
\def \s{~\rm{s}}
\def \km{~\rm{km}}

\def \K{~\rm{K}}
\def \g{~\rm{g}}

\def \AU{~\rm{AU}}

\def \yr{~\rm{yr}}

\def \days{~\rm{day}}

\def \mum{~\rm{\mu m}}

\date{\today}

\title{Planetary influences on photometric variation of the extreme helium subdwarf KIC~10449976}

\author{Ealeal Bear\altaffilmark{1} and Noam Soker\altaffilmark{1}}

\altaffiltext{1}{Department of Physics, Technion$-$Israel
Institute of Technology, Haifa 32000 Israel; ealealbh@gmail.com; soker@physics.technion.ac.il.}

\begin{document}

\begin{abstract}
We propose that the unstable $3.9 \days$ photometric periodicity of the hot subdwarf (sdO) KIC~10449976 results from a tidally locked planet that is heated to $\sim 5000 \K$ by the
UV radiation from the hot sdO star. Although the bolometric radiation from the planet is very small relative to that of the star, in the visible band
the planet contributes $\sim 0.07 \%$ of the light, sufficient to explain the observed periodic behavior.
In our proposed scenario the stochastic variations in period and light amplitude are attributed to weather on the planet.
Namely, streams on the surface and thermal variations in the planet's atmosphere that are driven by the heating and by the planet rotation lead to stochastic
changes in the amount of radiation emitted by the planets.
We predict that a careful monitoring will reveal a gas giant planet at an orbital separation of $8.3 R_\odot$ from KIC~10449976.
\end{abstract}

Keywords: KIC 10449976, stellar evolution, planet, stochastic.

\section{Introduction}
\label{sec:intro}

O and B subdwarfs (sdO, sdB) are hot and compact stars that burn helium in their core and have a thin
envelope (for the purpose of this paper we do not differentiate between them and we will refer to them as sdOB).
They are identified with Extreme Horizontal Branch (EHB) stars (e.g., \citealt{Heber1986, Heber2009}),
with stars evolving off-the zero age HB (ZAHB) through a `short-cut' from the EHB to the WD cooling track (so called AGB-Manque; \citealt{Dormanetal1993}),
or with the merger of two white dwarfs (WDs; e.g., \citealt{Webbink1984, Iben1990, Hanetal2003, Heber2008, Heber2009, Nelemans2010, Zhang2012}).
The later scenario is attractive when the star is hydrogen poor.
Most popular models for the formation of sdOB involve binary interaction (e.g., \citealt{Hanetal2003, Charpinetetal2008, Geieretal2010}).
These models are supported by the large fraction of sdB stars in binary systems (e.g., \citealt{Maxtedetal2001, Napiwotzkietal2004}).
In the first two scenarios, those of a star on the HB or evolving off the HB, the companion is required to remove mass from the red giant branch (RGB) progenitor (e.g., \citealt{Hanetal2007}).
Many of the binary systems have close orbits, implying that the system has gone through a common envelope (CE) phase during the
RGB phase of the sdOB progenitor (e.g., \citealt{Hanetal2002}).
The mass removal in principle can be done by a sub-stellar companion as well \citep{Soker1998, NelemansTauris1998, SokerHarpaz2000,
SokerHershenhorn2007, Politanoetal2008, VillaverLivio2007, VillaverLivio2009, Carlbergetal2009, BearSoker2010, Nordhausetal2010}.

In the case of a CE evolution of a sub-stellar companion to the RGB star, the question is whether
the sub-stellar companion survived the CE phase \citep{Bearetal2011, BearSoker2012, TutukovFedorova2013}.
\cite{BearSoker2012} suggest that the two Earth-size planets found by \cite{Charpinetetal2011}
around the sdB star KIC~05807616 are remnant of the tidally-destroyed metallic core of a massive planet.
Massive planets would even survive the CE phase, as in the case of the planet orbiting the metal-poor red HB star HIP~13044 (CD-36~1052) with an orbital
period of $P=16.2\pm 0.3$~days \citep{Setiawanetal2010}.
sdB stars with substellar companions at separations of $a_f \ga 1 \AU$ have also been found \citep{Silvottietal2007, Leetal2009, Qianetal2009, Qianetal2012}.
In these systems the wide substellar companions suggests the existence of a closer planet that went through the CE
phase and ejected the envelope of the RGB stellar progenitor of the sdOB star.
This closer planet would have been completely destroyed in the CE process.
\cite{Schuhetal2010} noted that the increasing number of substellar companions to sdB stars
may indicate the existence of an undiscovered planet population.

In this paper we examine whether the photometric periodic variation of the sdOB star KIC~10449976 can be accounted for by a planet companion.
It is an extreme helium-rich subdwarf that shows evidence for photometric periodic modulation of $P=3.9\days$ with an amplitude of
$\sim 0.02\%$ \citep{Jeffery2013}.
KIC 10449976 effective temperature is $T_{\rm eff}= 40000\pm 300^\circ \K$ and its surface gravity is $\log{g} = 5.3 \pm 0.1$.
\cite {Jeffery2013} argued that the sinusoidal modulation is probably an astrophysical variation and not of an instrumental origin,
and consider possible explanations as follows.
(1) Pulsation. The problem with pulsation is that for a star with the dimensions of KIC 10449976 the period of pulsations should
be of the order of $200\s \ll 3.9 \days$.
(2) Reflection from a close companion. \cite{Jeffery2013} could find a companion parameters space that can explain
the photometric amplitude and period, as well as the constrain that the radial velocity variability be $< 50 \pm 20 \km \s^{-1}$.
However, the lack of stability in the apparent period rules out a stable companion.
(3) Stellar spots. The photometric variation by stellar spots cannot be ruled out, but the  lack of stability in the
apparent period is problematic.

To address the unstable semi-periodic variation we examine a planet companion that suffers `weather instabilities'.
The advantage of a planet over a stellar companion is that the energy budget of the planet is dictated by radiation from the primary star, hence has a weather.
In section \ref{sec:Phys} we examine the parameter space of a possible planetary companion to KIC~10449976.
In section \ref{sec:Stochastic} we speculate on possible weather-processes that can account for the stochastic nature of this amplitude.
Our short summary is in section \ref{sec:Summary}.

\section{Physical properties of the proposed  KIC~10449976 planet}
\label{sec:Phys}

The effective temperature and surface gravity of KIC~10449976 are $T_{\rm eff}= 40000\pm 300 K$ and $\log{g}(\cm \s^{-2}) = 5.3 \pm 0.1$,
respectively \citep{Jeffery2013}.
Therefore the radius and luminosity of KIC~10449976 are
\begin{equation}
R_{\rm KIC} =  0.26 R_\odot \left( \frac{M_{\rm KIC}}{0.5M_\odot}\right)^{\frac{1}{2}}; \quad
L_{\rm KIC} = 156 \left( \frac{M_{\rm KIC}}{0.5M_\odot}\right) L_\odot,
\label{eq:RL}
\end{equation}
respectively, where $M_{\rm KIC}$ is the mass of KIC~10449976.
 The orbital separation of the assumed planet is calculated from the $P=3.9\days$ period of the photometric variation
\begin{equation}
a_{\rm p} =  8.3 \left(\frac{M}{0.5M_\odot}\right)^{\frac{1}{3}}  \left( \frac{P}{3.9\days}\right)^{\frac{2}{3}} R_\odot.
\label{eq:ap}
\end{equation}

Assuming a fraction $\eta$ of light absorption by the planet and emission as a black body from the half hemisphere facing the star (day side),
the surface temperature of the planet is
\begin{equation}
T_{\rm p}={T_{\rm KIC}} 2^{\frac{1}{2}}\frac{ {R_{\rm KIC}}^{\frac{1}{2}}}{(2 a_{\rm p})^{\frac{1}{2}}}
\eta^{\frac{1}{4}}
\simeq  5000 \left(\frac{M_{\rm KIC}}{0.5 M_\odot}\right)^\frac{1}{12} \left(\frac{T_{\rm KIC}}{40000^\circ K}\right)
\left( \frac{\eta}{0.5} \right)^{\frac{1}{4}} \K,
\label{eq:tp}
\end{equation}
where we used the period $P=3.9\days$ for the planet orbital period.

The ratio of bolometric luminosities between the planet when its hot (day) side faces us and the star for the above parameters is
\begin{equation}
\frac{L_{\rm KIC}}{L_{\rm p}}=3.6 \times 10^{-5} \left(\frac{83 R_{\rm p}}{a } \right)^{2}.
\label{eq:tL}
\end{equation}
This is a very small number, but the photometric modulation with an amplitude of $\sim 0.02 \%$ was detected with a photometer
on board the Kepler telescope in the spectral range of $0.4-0.85 \mum$.
For that spectral range we calculate the luminosities ratio $\chi$ when the hot planet hemisphere faces us to be
\begin{equation}
\chi \equiv \left( \frac{L_{\rm KIC}}{L_{\rm p}} \right)_{0.4-0.85\mum} \simeq 7 \times 10^{-4} .
\end{equation}
Such fluctuation amplitudes can explain the periodic modulation of $P=3.9\days$ with an amplitude of
$\sim 0.02\%$ \citep{Jeffery2013}.

The two advantages of a planet over a stellar companion is that the planet can account for the non-detection of radial velocity of the star, and that a planet can
have weather, hence accounting for the
stochastic variation. This will be studied in the next section, where we require that the planet spin is locked with the orbital motion.
We now examine whether the planet spin is indeed synchronized with the orbital motion.

From the circularization timescale $\tau_{\rm ep}$ given by \cite{Bodenheimeretal2001} we find the synchronization timescale of the planet
\begin{equation}
\tau_{\rm sp} \simeq \frac{I_{\rm p}}{M_{\rm p} R_{\rm p}^2}
\left(\frac{R_{\rm p}}{a} \right)^2 \tau_{\rm ep} \simeq 2 \times
10^4 \left( \frac{M_{\rm p}}{M_{\rm J}} \right) \left(
\frac{M_{\rm KIC}}{0.5M_\odot} \right)^{-\frac{3}{2}} \left( \frac{R_{\rm
p}}{R_{\rm J}} \right)^{-3} \left( \frac{a}{8.3 R_\odot}
\right)^{\frac{9}{2}} \yr,
\label{eq:Sync1}
\end{equation}
where $I_{\rm p}\simeq 0.3 M_{\rm p} R_{\rm p}^2$ is the moment of inertia of the planet.
Despite the large uncertainties in the tidal mechanism strength, this time is much shorter than the HB life time of $\sim 10^8 \yr$, and
we can safely conclude that such a planet is synchronized with the orbital motion.

Another process that can manifest changes in the atmosphere is gas evaporation.
Evaporation is observed for planets around MS stars. For example, the planet $WASP-12b$ orbiting a MS star with an orbital period of $P= 1.09 \days$
is presumed to be losing mass \citep{Cowan2012}.
\cite{BearSoker2011b} previously derived an expression for evaporation of planets around sdOB stars similar to KIC~10449976 (eq. 12 and Fig 2 there).
Scaling to KIC~10449976 and a planet of mass $M_p=15M_J$, we find the mass evaporation rate from the assumed planet around
KIC~10449976 to be $\dot M_p \sim 10^{14} \g \s^{-1}$ (the mass-loss rate is defined positively).
We note that \cite{Owen2012} get a similar range of mass-loss rates $\dot M_p \simeq 5\times 10^{13} - 10^{14} \g \s^{-1}$ for
an orbital separation of $a \simeq 0.1\AU$, and a planet mass of  $M_p \sim 1 M_J$ orbiting a solar type star
(for more details see Fig 9 in \citealt{Owen2012}).
Within the life time of an HB star, $\sim 1 M_{\rm J}$ will be evaporated.
Such an intensive mass-loss process might contribute to dynamical changes on the surface of the planet.

\section{Stochastic variations}
\label{sec:Stochastic}

In our scenario of a hot Jupiter we attribute the stochastic variations to weather on the planet.
We will consider streams in the planet and thermal variations. The discussion in this section is of a more speculative nature,
as the processes require deeper studies. However, the processes do take place in solar system planets,
and are expected to be more vigourous in hot Jupiters.
The study of the causes of the weather variations is beyond the scope of our paper. We limit ourselves to show that
the typical timescale of these processes is of the order of the orbital period, and hence can cause unstable
periodic behavior.

\subsection{The stream timescale}
Flows in the atmosphere of hot Jupiters were discussed before (e.g., \citealt{Showman2013, Showman2011}).
The streams exchange hot and cold atmosphere gas between the day and night half-hemispheres.
\cite{Spiegel2012} state that the typical velocity for streams (winds) in hot Jupiters is $\sim 1 \km \s^{-1}$.
This implies a typical timescale of
\begin{equation}
\tau_{\rm stream} \sim \frac{\pi R_{\rm p}}{v_{\rm stream}} = 2.5 \left(\frac{R_{\rm p}}{R_{\rm J}}\right)
\left(\frac{v_{\rm stream}}{1 \km \s^{-1}}\right)^{-1}\days.
\label{eq:tstream}
\end{equation}

Another important feature for our proposed scenario is that the zonal bands of the streams (jets) in hot Jupiters are wide \citep{Spiegel2012}.
This implies that each stream covers a large fraction of the surface.
Let us consider a demonstrative arbitrary example.
If a stream brings cooler gas to the day side of a temperature of
$3000 \K$, in an area that covers $10 \%$ of the day side time, the planet luminosity will drop by $9 \%$.
We note that the plant's temperature here is much larger than that of hot Jupiters around MS stars, and the stream
influence is expected to be larger than that in typical hot Jupiters.

The stream timescale of $\sim 2 \days$ is of the order of the orbital period of $3.9 \days$.
Hence variations due to the streams, if exist, can change the time between two consecutive maxima of plant emission,
explaining the non stable period.

\subsection{The thermal timescale}
Another possible source of luminosity fluctuations can be thermal changes in the photosphere.
If the gas suffers thermal variations this will change the energy radiated.
The exact source of thermal variations, e.g., thermal instabilities or fluctuations driven by unstable evaporation,
will be studied in a separate paper. To calculate the thermal timescale we estimate the pressure scale height
\begin{equation}
H_{\rm P} =\left(\frac{GM_{\rm p}}{R_{\rm p}^2}\right)^{-1} \frac{kT_{\rm p}}{\mu m_h}\sim
0.002 \left(\frac{M_{\rm p}}{10M_{\rm J}}\right)^{-1}\left(\frac{R_{\rm p}}{1R_{\rm J}}\right)^2
\left(\frac{T_{\rm p}}{5000\K}\right) R_{\rm J},
\end{equation}
where for the molecular weight we take $\mu=1$, and the other symbols have their usual meaning.
The density at the photosphere is given by (e.g., \citealt{Kippenhahn1990})
\begin{equation}
\rho_{\rm p} = 1.4\times 10^{-6}\left(\frac{M_{\rm p}}{10 M_{\rm J}}\right)\left(\frac{\kappa}{0.03 \cm^{2}\g^{-1}}\right)^{-1}
\left(\frac{R_{\rm p}}{1R_{\rm J}}\right)^{-2}
\left(\frac{T_{\rm p}}{5000 \K}\right) ^{-1}\g \cm^{-3},
\end{equation}
where the opacity is $\kappa \sim 0.03 \times 10^{-2} \cm^{2}\g^{-1}$ for $T_{\rm
eff}=5000\K $ \citep{Alexander1994}.
We realize that this formula applies for stars. However, for our first order approximation of hot Jupiter planets
which have high luminosity and a temperature as stars this is a reasonable approximation.

Taking the mass in the photosphere (day side) $M_{\rm photos} \simeq 2 \pi R_{\rm }^2 \rho_{\rm p} H_{\rm P}$, we find
\begin{equation}
M_{\rm photos} \sim 10^{22}\left(\frac{R_{\rm p}}{R_{\rm J}}\right)^2\left(\frac{\kappa}{0.03 \cm^{2}\g^{-1} }\right)^{-1} \g.
\end{equation}

Calculating the thermal timescale for the planet:
\begin{equation}
\begin{split}
\tau_{\rm thermal} \simeq \frac{GM_{\rm p}\times M_{\rm photos}}{2R_{\rm p} L_{\rm abs}}
\newline
\sim 1  \left(\frac{\eta}{0.5}\right)^{-1}
\left(\frac{M_{\rm p}}{10M_{\rm J}}\right)
\left(\frac{\kappa}{0.03 \cm^{2}\g^{-1} }\right)^{-1}
                 \\  \times
\left(\frac{R_{\rm p}}{R_{\rm J}}\right)^{-1}
\left(\frac{a_{\rm p}}{8.3 R_\odot }\right)^{2}
\left(\frac{L_{\rm KIC}}{156 L_\odot }\right)^{-1}\days,
\end{split}
\label{eq:tthermal}
\end{equation}
where $L_{\rm abs}$ is the luminosity absorbed by the planet.
The thermal timescale is of the same order as the orbital period, $3.9 \days$,
and can cause some of the stochastic behavior
of the light curve.  However, this is just a rough approximation, further investigation of this subject is needed.

\section{Summary}
\label{sec:Summary}

In a recent paper \citep{Jeffery2013} report a $3.9 \days$ periodic photometric variation in the hot subdwarf star KIC~10449976.
The period is not stable, neither in time nor in amplitude.
\cite{Jeffery2013} discuss three possible scenarios as possible explanations: stelar spots, pulsation, and their favorable
explanation of reflection from a "stellar" companion.
However, they do not account for the unstable periodic behavior.

In this paper we offered an alternative scenario for the unstable periodic behavior of KIC~10449976 based on
a planet (or a low-mass brown dwarf) that orbits KIC~10449976 with an orbital period of $3.9 \days$, implying an orbital
separation of $a_p=8.3 R_\odot$ (section \ref{sec:Phys}).
We proposed that the UV radiation from KIC~10449976, whose effective temperature is $T_{\rm eff}= 40000\pm 300 K$, heats a tidally locked planet
to a temperature of $\sim 5000 \K$ (eq. \ref{eq:tp}).
The radiation from the hot planet causes the observed amplitude variation.
The advantages of a planet over a stellar companion is that a low mass planet can account for the very low orbital motion of the star (undetected),
and that a planet has a weather because the energy supplied by the star is much larger than the internal energy supplied by the planet, and the
planet spin is locked to the orbital motion. These ensure steep temperature gradients between day and night sides that drive the weather.

In section \ref{sec:Stochastic} we suggested that the weather on the planet accounts for the stochastic variation observed in the amplitude
and periodic variation.
The stochastic behavior of the weather includes streams of gas on the surface of the planet and thermal changes that modulate
the effective temperature of the photosphere.
These changes in effective temperature are responsible for the unstable amplitude (section \ref{sec:Stochastic}).
We showed that if streams and thermal instability exist, the typical timescale for their variation is $\sim 1-3 \days$
(eqs. \ref{eq:tstream} and \ref{eq:tthermal}), not much lower than the orbital period. Therefore, the amplitude modulation
due to stream and thermal instability may account for the unstable periodicity of KIC~10449976.

The planetary scenario proposed here predicts that the orbital velocity of KIC~10449976 has an amplitude of
$\sim 1 \km \s^{-1}$.

We thank the anonymous referee for significant comments that substantially improved our planetary scenario.
This research was supported by the Asher Fund for Space Research at the Technion.
%

\end{document}